\documentclass[twocolumn]{jpsj2} 
\newcommand{\Cel}{\ensuremath{C_{\rm el}}}
\newcommand{\Tco}{\ensuremath{T_{\rm co}}}
\newcommand{\kB}{\ensuremath{k_{\rm B}}}
\newcommand{\Tc}{\ensuremath{T_{\rm c}}}
\newcommand{\Deff}{\ensuremath{\Delta _0^{\rm eff}}}
\newcommand{\EF}{\ensuremath{E_{\rm F}}}
\newcommand{\Ctotal}{\ensuremath{C_{\rm total}}}
\newcommand{\Cph}{\ensuremath{C_{\rm ph}}}
\newcommand{\Td}{\ensuremath{T_{\rm d}}}
\newcommand{\Sn}{\ensuremath{S_{\rm n}}}
\newcommand{\Ss}{\ensuremath{S_{\rm s}}}
\newcommand{\gn}{\ensuremath{\gamma _{\rm n}}}
\newcommand{\gs}{\ensuremath{\gamma _{\rm s}}}
\newcommand{\Tsf}{\ensuremath{T_{\rm sf}}}
\newcommand{\go}{\ensuremath{\gamma _{\rm o}}}
\newcommand{\Tcmax}{\ensuremath{T_{\rm c}^{\rm max}}}
\newcommand{\pmax}{\ensuremath{p_{\rm m}}}
\newcommand{\Um}{\ensuremath{U_{\rm m}}}
\newcommand{\Dcoh}{\ensuremath{\Delta _{\rm coh}}}
\newcommand{\TonNernst}{\ensuremath{T_{\rm on}^{\rm Nernst}}}

\title{Electronic Specific Heat of \textbf{La}$_{2-x}$\textbf{Sr}$_{x}$\textbf{CuO}$_{4}$: \\Pseudogap Formation and 
Reduction \\of the Superconducting Condensation Energy}

\author{Toshiaki \textsc{Matsuzaki}
, Naoki \textsc{Momono}
, Migaku \textsc{Oda}
, and Masayuki \textsc{Ido}
}

\inst{
Department of Physics, Hokkaido University, Sapporo 060-0810, Japan }

\abst{To examine the so-called small pseudogap and the superconducting (SC) 
condensation energy $U(0)$, the electronic specific heat $C_{el}$ was measured 
on La$_{2 - x}$Sr$_{x}$CuO$_{4}$ up to $\sim 120$ K. In samples with doping 
level $p (=x)$ less than $\sim 0.2$, small pseudogap behavior appears in the 
$\gamma \ (=\Cel /T)$ vs. $T$ curve around the mean-field critical temperature 
for a $d$-wave superconductor$ \Tco \ (=2\Delta _{0}/(4\sim 5)\kB )$, 
where $\Delta _{0}$ is the maximum gap at $T \ll \Tc $. The condensation 
energy $U(0)$ is largely reduced in the pseudogap regime ($p
\lesssim  0.2$). The reduction of $U(0)$ can 
be well reproduced by introducing an effective SC energy scale $\Deff =\beta p\Delta _{0} \ (\beta =4.5)$ instead of $\Delta 
_{0}$. The effective SC energy scale is discussed in relation to the 
coherent pairing gap formed over the nodal Fermi arc. 
}

\kword{electronic specific heat, superconducting condensation energy, 
pseudogap, La$_{2 - x}$Sr$_{x}$Cu$_{x}$O$_{4}$}

\begin{document}
\maketitle

\section{Introduction} 

Loram \textit{et al}. revealed that the superconducting (SC) condensation energy $U(0)$ of 
high-$\Tc $ cuprates was exceedingly suppressed even in a slightly 
underdoped region, where $\Tc $ was still high enough\cite{Loram1,Loram2,Loram3}. To clarify 
the origin of the exceeding reduction of $U(0)$ is expected to give an 
important clue to understanding the mechanism of high-$\Tc $ 
superconductivity, because the condensation energy $U(0)$ reflects features of 
the pairing mechanism and/or the collective motion of pairs. Thus this 
problem has been discussed from various points of view\cite{Loram1,Loram2,Loram3,Demler1,Anderson1,Lee1}. Loram \textit{et al}. 
have argued the exceeding reduction of $U(0)$ in terms of the existence of a 
$T$-independent energy gap at the Fermi level $\EF $, which is persistent up to 
a slight overdoping level\cite{Loram1,Loram2,Loram3}. Another explanation has been proposed 
by Demler and Zhang on the basis of a spin-triplet particle-particle 
resonance, which lowers the antiferromagnetic (AFM) exchange energy in the 
$d$-wave SC state through the formation of the so-called $\pi $-resonance in 
the dynamic spin susceptibility $\chi (q,\omega )$\cite{Demler1,Scalapino1}. In this 
scenario, the SC condensation energy $U(0)$ comes from the difference of AFM 
exchange energy between the normal and superconducting states, and so $U(0)$ 
will be reduced greatly in samples with small doping levels whose AFM 
correlation is already enhanced to a large extent in the normal state. Other 
kind of explanation is that in the underdoped region the superconductivity 
will be driven by a small gain in kinetic energy caused by interplane phase 
coherence, which naturally leads to a small $U(0)$\cite{Anderson1}. Recently Lee and 
Salk have claimed that unusual doping-level ($p$) dependence of $U(0)$ can be 
explained within the framework of the boson-pair condensation in the SU(\ref{eq2}) 
slave-boson model\cite{Lee1}.

Very recently it was demonstrated for La$_{2 - x}$Sr$_{x}$CuO$_{4}$ (La214) 
that the exceeding reduction of $U(0)$ can be reproduced quantitatively by 
adopting the SC energy scale $\beta p\Delta _{0} \ (\beta =4.5)$ instead 
of $\Delta _{0}$, where $\Delta _{0}$ is the maximum value of the 
$d$-wave energy gap\cite{Momono3}. The SC energy scale $\beta p\Delta _{0} \ (\beta 
=4.5)$, deduced from the phenomenological relation $\Tc \sim \kappa 
p\Delta _{0} \ (\kappa \sim 1.7)$\cite{Ido1}, becomes much smaller than 
$\Delta _{0}$ at small doping levels through the factor $\beta p$, and 
$U(0)$ will be markedly suppressed there. On the other hand, the energy scale 
$\beta p\Delta _{0} \ (\beta =4.5)$ becomes comparable with $\Delta 
_{0}$ around $p\ (=x)=0.22$, where no pseudogap behavior appears and the SC 
properties are of the BCS type. This leads naturally to the speculation that 
the change of the SC energy scale from $\Delta _{0}$ to $\beta p\Delta 
_{0}$ will result from the development of a pseudogap in the normal state. 

In many high-$\Tc $ cuprates, two types of pseudogaps appear in the normal 
state for $x \lesssim  0.2$; one is the so-called large 
pseudogap brought by the downward shift of flat bands from $\EF $ near $(\pi 
,0)$ and $(0,\pi )$, and the other is the so-called small pseudogap 
characterized by an energy scale of the order of $\Delta _{0}$\cite{Ino1,Ino2,Takahashi2,Loeser1,Harris1,Renner2,Timusk1}.
The small pseudogap was first reported as the spin gap in NMR 
relaxation time $T_{1}$ measurements on YBa$_{2}$Cu$_{3}$O$_{6 + \delta 
}$\cite{Yasuoka1}, and has been found in many high-$\Tc $ cuprates although it was 
as late as last year that the spin gap was reported in inelastic neutron 
scattering and NMR-$T_{1}$ measurements on La214\cite{Lee2,Itoh2}. Angle-resolved 
photoemission spectroscopy (ARPES) measurements on 
Bi$_{2}$Sr$_{2}$CaCu$_{2}$O$_{8 + \delta }$ (Bi2212) have clarified that the 
small pseudogap starts to open at the Fermi surface near ($\pi $,0) and 
(0,$\pi )$ at temperature $T^{\ast }(>\Tc )$. The small pseudogap grows 
towards the $d$-wave nodal points near $(\pi /2,\pi /2)$ at 
$\Tc <T<T^{\ast }$, leaving the so-called nodal Fermi arc centered at the 
nodal points\cite{Norman1}. Very recently Yoshida \textit{et al}. and Zhou \textit{et al}. reported the 
existence of the nodal Fermi arc in underdoped samples of La214\cite{Yoshida2,Zhou1}. 
Tunneling spectroscopy measurements on Bi2212 have demonstrated that the 
small pseudogap behavior becomes evident gradually around the mean-field 
critical temperature $\Tco =2\Delta _{0}/(4\sim 5)\kB $ for a 
$d$-wave superconductor\cite{Nakano2}. 

In the present study, the electronic specific heat $\Cel $ for La214 was 
systematically measured over a wide doping-level $(p)$ range to examine the 
pseudogap formation, the marked reduction of $U(0)$ and the interrelation 
between them in detail. The small pseudogap behavior appears around the 
temperature $T'$ $(\sim \Tco )$ which roughly correlates with the onset 
temperature of the enhanced Nernst signal reported by Wang \textit{et al}\cite{Wang1}. It was 
reconfirmed that the reduction of $U(0)$ becomes more conspicuous in samples 
with higher $T'$ and can be well explained by the SC energy scale $\beta 
p\Delta _{0}$ $(\beta =4.5)$ over a wide $p$ range. The SC energy scale is 
discussed in relation to the shrinkage of the coherent part of the pairing 
gap in the pseudogap regime. 

\section{Experimental}

Ceramic samples of La214 used for the present study were prepared by using a 
solid reaction in an oxygen atmosphere. The SC critical temperature 
$\Tc $ was determined from the SC diamagnetism measured with a SQUID 
magnetometer. Specific heat measurements were carried out using a 
conventional pulsed-heat technique. 

The electronic specific heat $\Cel $ of SC samples was obtained by 
subtracting the phonon term $\Cph $ of an impurity-doped nonsuperconducting 
sample from the observed total specific heat $\Ctotal $; 
$\Cel =\Ctotal -\Cph $. The phonon term $\Cph $ of the 
nonsuperconducting sample was obtained as follows. First we determined the 
coefficient of the $T$ linear term of $\Cel $, $\gamma $, using a 
$\Cel $/$T$ vs. $T^{2}$ plot at low temperatures where $\Cph $ shows $T^{3}$ 
dependence\cite{Momono7}. Next we extracted $\Cph $ by subtracting the electronic 
term $\gamma T$ from $\Ctotal $ on the assumption that $\gamma $ was 
independent of $T$. We tried to use Zn impurity at first to suppress the 
superconductivity in the process of obtaining $\Cph $, because the Zn ion 
has the mass closest to that of Cu$^{2 + }$ and carries no local magnetic 
moment. However, since the Zn impurity modifies the phonon properties, the 
phonon term $\Cph ^{\rm Zn}$ of the Zn-doped non-superconducting sample 
becomes appreciably different from the $\Cph $ of the SC sample, as will be 
described in the following section. Then we used the phonon term 
$\Cph ^{\rm Ni}$ obtained for an Ni-doped non-superconducting sample, although 
the Ni impurity carries local magnetic moment. It has been revealed that a 
small amount of Ni impurity removed the superconductivity and the small 
pseudogap behavior\cite{Nagata1}.

\section{Results and Discussion}

\subsection{Electronic Specific Heat $\Cel $ of ${\rm La}_{2 - x}{\rm Sr}_{x}{\rm CuO}_{4}$}

\begin{figure}[htbp]
\begin{center}
\includegraphics[width=0.95\linewidth,clip]{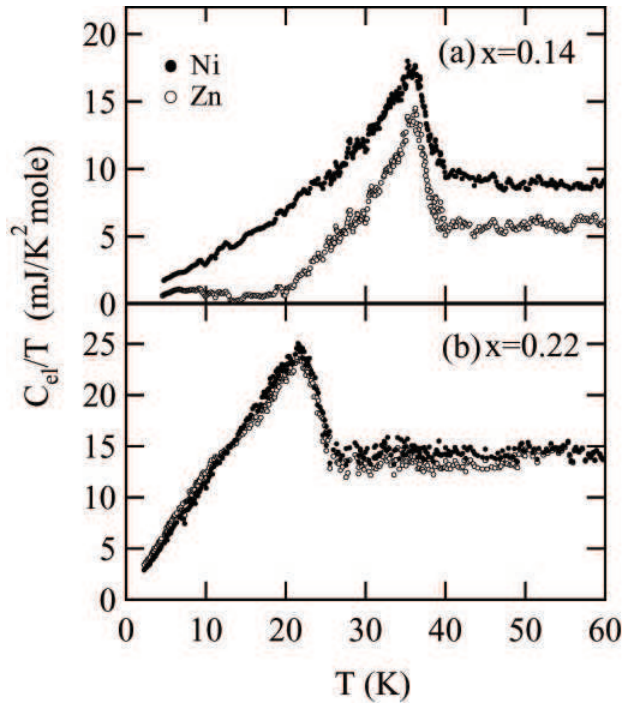}
\end{center}
\caption{$T$-dependence of the electronic specific heat $\Cel $ ($\Cel $/$T$ vs. 
$T$ plot) obtained for La$_{2 - x}$Sr$_{x}$CuO$_{4}$ with $x=0.14$ and 0.22 by 
using different phonon terms $\Cph ^{\rm Zn}(\circ )$ and 
$\Cph ^{\rm Ni}(\bullet)$. }
\label{f1}
\end{figure}

In Fig. \ref{f1}
, typical temperature dependences of $\Cel $, obtained by using the 
phonon terms$ \Cph ^{\rm Ni}$ and $\Cph ^{\rm Zn}$, are shown with a 
$\Cel /T$ $(=\gamma )$ vs. $T$ plot. The $\gamma -T$ plot for $\Cph ^{\rm Zn}$ shows 
a seeming anomaly around 15 K, and is severely distorted over the 
temperature range examined. On the other hand, the $\gamma -T$ plot for$ \Cph ^{\rm Ni}$ 
shows no anomaly around 15 K, and shows a plausible $T$-dependence of $\gamma $ 
below and above $\Tc $. These results imply that Zn impurity will seriously 
change the phonon term $\Cph $ whereas the influence of Ni is very small. It 
has also been reported for La214 that the Zn- and Ni-impurity effects on the 
structural phase transition from the tetragonal phase to the orthorhombic 
one at temperature $\Td  (\gg \Tc )$ are contrasting with each; $\Td $ is 
little influenced by doping with Ni whereas it is enhanced appreciably by 
doping with Zn\cite{Momono4}. These facts mean that the nature of the phonon 
system is largely modified by doping with Zn.

In high $\Tc $ cuprates, it has been clarified that Zn impurity with no 
3$d$-spin disturbs the AFM Cu-3$d$-spin correlation around Zn more seriously than 
Ni-impurity with 3$d$-spin\cite{Ishida1,Nakano3,Julien1,Itoh1}. Since the AFM 3$d$-spin correlation 
couples with B$_{2\rm u}$ phonon modes, as has been reported in neutron 
scattering experiments on high-$\Tc $ cuprates\cite{Harashina1}, we can conjecture 
that the nature of the phonon system will be significantly modified by the 
serious disturbance of 3$d$-spin correlation caused by doping with Zn. This 
conjecture is consistent with the fact that the Zn-impurity effect on 
$\Cph $ becomes less evident in the highly-doped $x=0.22$ sample, where the AFM 
spin correlation is weakened to a large extent, as seen in Fig. \ref{f1}
.

\subsection{Superconducting Anomaly and Pseudogap Behavior in $\Cel $}

Figure 2 shows the $p$ dependence of $\Cel $, obtained by subtracting the 
phonon term $\Cph ^{\rm Ni}$ from $\Ctotal $, with a $\Cel /T$ $(=\gamma )$ vs. 
$T$ plot. The SC anomaly appears clearly in the $\gamma $ vs. $T$ plot for all 
samples investigated. The anomaly for $x=0.22$ is very similar, in both shape 
and size, to the BCS result for a $d$-wave superconductor over a wide $T$ range 
except just below and above $\Tc $, where the SC critical fluctuation 
effects become evident. On the other hand, the anomaly for $x<0.2$ becomes 
rather different from the BCS result; namely, the $\gamma $ value reaches 
the peak value at $T \cong T_{c }$and tends to decrease more rapidly at 
$T<\Tc $. In particular, the $\gamma $ value for $x \le 0.1$ decreases very 
steeply at $T<\Tc $, leaving a sharp peak at $\Tc $. The rapid decrease of 
$\gamma $ at $T<\Tc $ implies that the energy gap is developed to a large 
extent just below $\Tc $, as observed in tunneling spectroscopy 
measurements\cite{Renner2,Dipasupil1}, and quasiparticle excitations are rather suppressed 
even in the neighborhood of $\Tc $. On the other hand, the $\gamma $ value 
tends to be enhanced at temperatures just above $\Tc $, in particular, in 
samples with small doping levels. Some inhomogeneity effect, leading to the 
distribution of $\Tc $, could cause the enhancement of $\gamma $ at 
$T>\Tc $. In this case, the peak anomaly at $\Tc $ would be rounded owing to 
the distribution of $\Tc $. However, this is not the present case, because 
the peak anomaly at $\Tc $ is very sharp even in $x=0.08$ and 0.1 samples whose 
$\gamma $ enhancement at $T>\Tc $ is most evident among samples examined. 
This fact means that the enhancement of $\gamma $ at $T >\Tc $ is not due to 
an inhomogeneous effect but due to the strong SC critical fluctuation 
effect.
\begin{figure}[tb]
\begin{center}
\includegraphics[width=0.95\linewidth,clip]{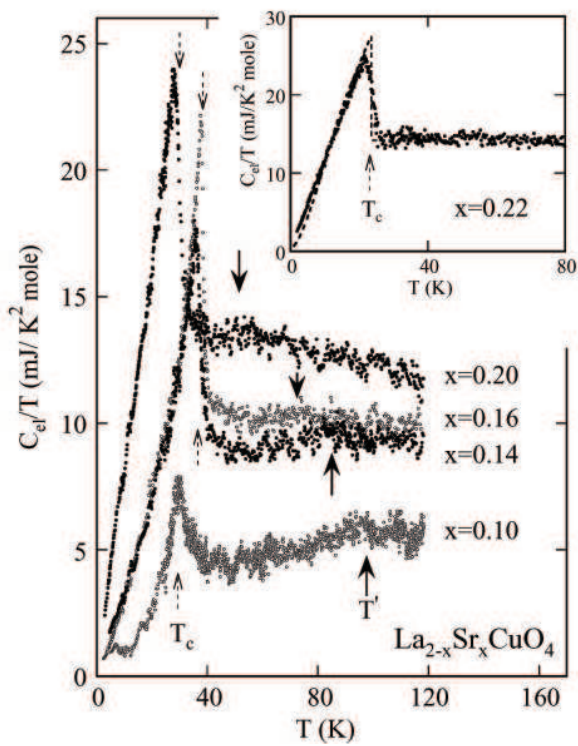}
\end{center}
\caption{Electronic specific heat $\Cel $ for La$_{2 - x}$Sr$_{x}$CuO$_{4}$ 
($0.1 \le x \le  0.2$) plotted with $\Cel /T$ vs. $T$. The inset shows the 
$\Cel /T$ vs. $T$ curve for $x=0.22$. The dotted line in the inset represents the 
theoretical result for a $d$-wave BCS superconductor.}
\label{f2}
\end{figure}

\begin{figure}[htbp]
\begin{center}
\includegraphics[width=0.95\linewidth,clip]{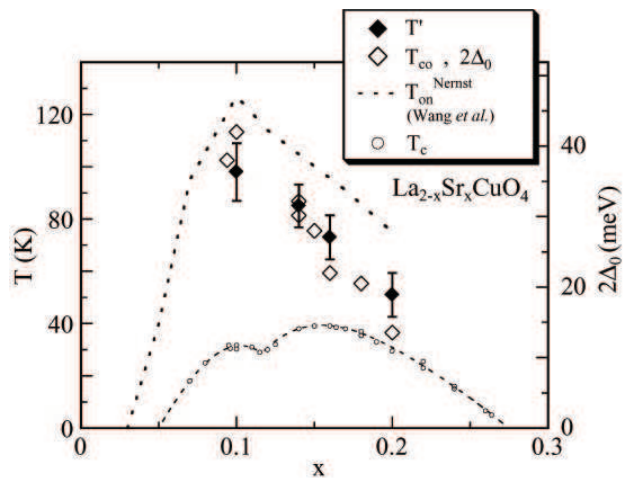}
\end{center}
\caption{Doping-level dependence of $T'$, $\Tco $, 2$\Delta _{0}$ and 
$\Tc $. Onset temperature of the enhanced Nernst signal 
($\TonNernst $) is also shown by the dotted line.\cite{Wang1} Note that for 
the open rhombus the left and right axes provide the scales of $\Tco $ and 
2$\Delta _{0}$, respectively.}
\label{f3}
\end{figure}

It should be noted here that the $\gamma -T$ curve for $x 
\lesssim  $0.20 shows a small, broad bump 
at a certain temperature $T'$ $(>\Tc )$ in the normal state, and the $\gamma $ 
value gradually decreases at $T<T'$. The temperature $T'$ increases with lowering $p$ $(=x)$, 
and the decrease of $\gamma $ at $T<T'$ becomes evident. The decrease of $\gamma 
$ at $T<T'$ implies that a small pseudogap around the Fermi energy $\EF $ will 
start to evolve around $T'$ and suppress the density of states (DOS) at 
$\EF $, $N(0)$. As has been reported, $T'$ is in agreement with the mean-field 
critical temperature $\Tco \sim 2\Delta _{0}/4.3\kB $ (Fig. \ref{f3}
)\cite{Ohkawa,Won1,Tanamoto1}, where the maximum gap $\Delta _{0}$ was measured in tunneling 
spectroscopy experiments on La214\cite{Nakano2,Oda1,Matsuzaki1}. The in-plane 
resistivity and the uniform susceptibility of La214 also show small 
anomalies around $\Tco (\sim T')$, as in Bi2212\cite{Nakano2}. 

Recently, it was reported for La214 and Bi$_{2}$Sr$_{2 - 
y}$La$_{y}$CuO$_{6}$ that the Nernst signal enhanced by vortices or 
vortex-like excitations extends to temperatures well above 
$\Tc $\cite{Wang1}. The anomalous Nernst signal has been interpreted in terms of 
strong fluctuations between the pseudogap state and $d$-wave SC condensates. It 
is worthwhile to point out the fact that both the small, broad bump in the 
$\gamma -T$ curve and the enhanced Nernst signal appear in La214 samples for 
$x \lesssim  0.2$, and the temperature $T'$ ($\sim 
\Tco )$ exhibiting the bump in the $\gamma -T$ curve roughly correlates 
with the onset temperature of the enhanced Nernst signal 
$\TonNernst $, as shown in Fig. \ref{f3}
. Such a correlation between $T'$ and 
$\TonNernst $ confirms that the anomalous Nernst signal will be 
intimately related to the development of the small pseudogap in the normal 
state. 

\subsection{Superconducting Condensation Energy of ${\rm La}_{2 - x}{\rm Sr}_{x}{\rm CuO}_{4}$}\label{3.3}

The SC condensation energy $U(0)$ can be evaluated by integrating the entropy 
difference $\Sn -\Ss $ between $T=0$ and $\Tc $
\begin{equation}
\label{eq1}
U(0) = \int_{0}^{\Tc  } (\Sn  - \Ss )dT
,
\end{equation}

\noindent
where the subscripts s and n stand for the SC and hypothetical normal states 
at $T<\Tc $, respectively. Given both $\gs $ and $\gn $ as 
a function of $T$, we can obtain the entropy $\Ss $ and $\Sn $ by executing the 
integration 
\[
S_{s,n} (T) = \int_{0}^{T} {\gamma _{s,n} 
dT} ,
\]

\noindent
and evaluate the condensation energy $U(0)$ using eq. (\ref{eq1}). In the present 
system, the SC critical fluctuation effect is so strong that we have to take 
the upper limit of the integration of eq. (\ref{eq1}) to be $\Tsf $ $(>\Tc )$, 
instead of $\Tc $, where $\gamma _{n}$ $(T>\Tc )$ begins to increase on 
account of the SC critical fluctuation effect. 
\begin{figure}[htbp]
\begin{center}
\includegraphics[width=0.95\linewidth,clip]{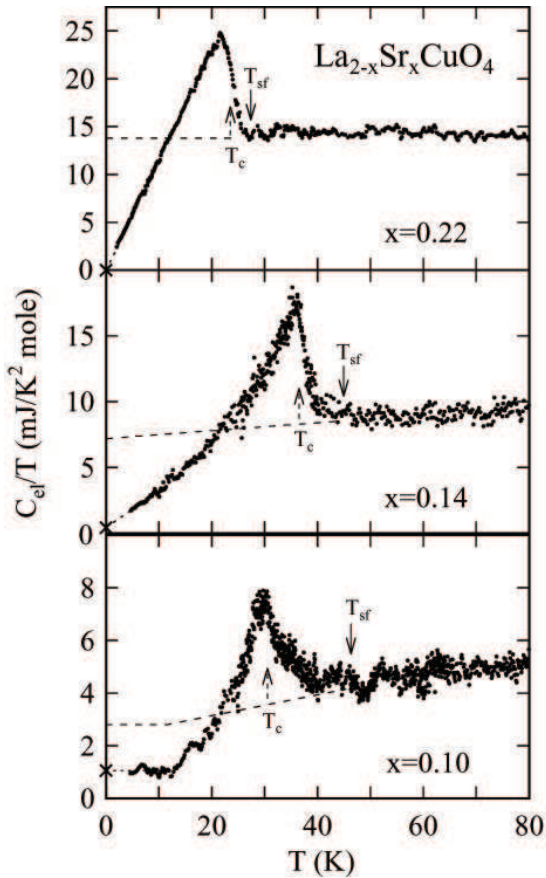}
\end{center}
\caption{$T$-dependence of $\Cel /T$ ($=\gamma $) in the superconducting state, 
$\gs (T)$, and the normal state, $\gn (T)$, for La$_{2 - 
x}$Sr$_{x}$CuO$_{4}$. The broken line represents the hypothetical normal 
state value $\gn (T)$ at $T<\Tc $. The residual $\gamma $-value at 
$T=0$ ($\go )$ in the superconducting state, which was estimated in 
$\Cel /T$ vs. $T^{2}$ plots at low temperatures, is also shown on the vertical 
axis ($\times )$.}
\label{f4}
\end{figure}

\begin{figure}[htbp]
\begin{center}
\includegraphics[width=0.95\linewidth,clip]{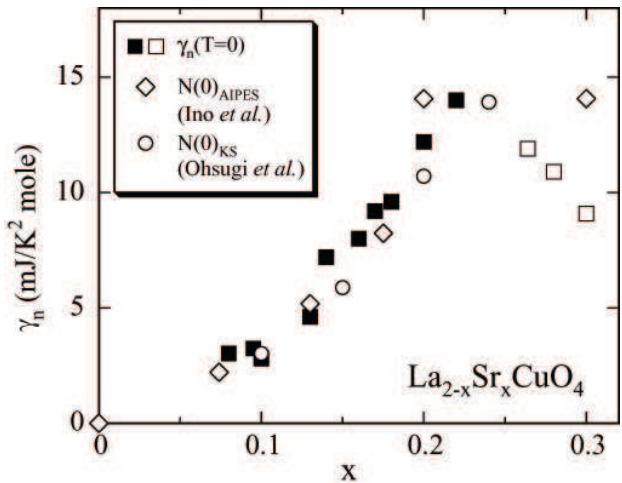}
\end{center}
\caption{Doping-level dependence of $\gn (0)$ for La$_{2 - 
x}$Sr$_{x}$CuO$_{4}$. The $\gn (0)$ for non-superconducting samples 
with $x>0.26$ were obtained from the conventional $C/T$ vs. $T^{2}$ plots. The open 
rhombus represents the density of states at $\EF $, $N(0)_{\rm AIPES}$, reported 
for angle-integrated photoemission spectroscopy measurements on La$_{2 - 
x}$Sr$_{x}$CuO$_{4}$.\cite{Ino1} The open circle represents $N(0)_{\rm KS}$ 
estimated from the drop of the Knight shift below $\Tc $.\cite{Ohsugi1,Matsuzaki2}
The $N(0)_{\rm AIPES}$ and $N(0)_{\rm KS}$ are normalized with $\gn (0)$ at 
$x \approx 0.22$.}
\label{f5}
\end{figure}

\begin{figure}[htbp]
\begin{center}
\includegraphics[width=0.95\linewidth,clip]{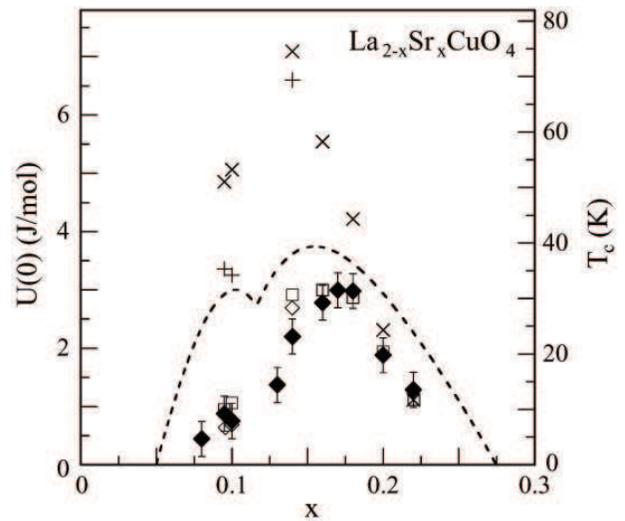}
\end{center}
\caption{Superconducting condensation energy $U(0)$ for La$_{2 - 
x}$Sr$_{x}$CuO$_{4}$. 

Experimental data ($\blacklozenge $) obtained by executing the integral of eq. (\ref{eq1}) are 
shown, together with the calculated values ($\times $) given by eq. (\ref{eq2}) for 
the experimental values of $\gn (0)$ and $\Delta _{0}$. The 
$U(0)$ ($\square $) calculated by substituting $\Deff $ ($=\beta 
p\Delta _{0}$) for $\Delta _{0}$ in eq. (\ref{eq2}) are also shown. The 
condensation energy $U(0)$ was also calculated for $\Delta _{0}$(+) and 
$\Deff (\lozenge )$ using $\gn ' = \gn (0) -\go $ instead of $\gn $(0). The dotted line 
shows the $p$ dependence of $\Tc $.}
\label{f6}
\end{figure}

In the present study, to determine the $T$ dependence of $\gamma _{n }$for 
the hypothetical normal state, we extrapolated the data of $\gn $ 
measured at $T>\Tsf $ down to below $\Tsf $, as shown in Fig. \ref{f4}
. In the 
highly-doped $x=0.22$ sample, since $\gn $ is almost constant at 
$T>\Tsf $, we can safely extrapolate the data down to $T=0$. The hypothetical 
normal value $\gn (T<\Tsf )$ thus obtained satisfies the 
constraint on the second order phase transition; namely, 
$\Sn (\Tsf )=\Ss (\Tsf )$. This constraint is often called ``the 
entropy balance'', because excess and deficit areas of the $\gs $ 
vs. $T$ curve, compared with $\gn $ $(T<\Tsf )$, must be equal with each 
other to satisfy the condition $\Sn (\Tsf )=\Ss (\Tsf )$. On the other 
hand, the $\gn $-value for $x 
\lesssim  0.2$ becomes temperature 
dependent at $T>\Tsf $ on account of the small pseudogap formation, so we 
extrapolated the data of $\gn (T>\Tsf )$ down to below $\Tsf $ 
using a declining straight line to satisfy the entropy balance (Fig. \ref{f4}
). 
Furthermore, in samples for $x \le 0.1$, we took $\gn $
 $(T<\Tsf )$ to be constant at $T<0.3\Tc $ because $\gs $ tends 
to be saturated at $T<0.3\Tc $ in these samples. In Fig. \ref{f5}
, the extrapolated 
value of $\gn$ $(T<\Tsf )$ at $T=0$, $\gn (0)$, is shown 
against $p$ $(=x)$. The $p$ dependence of $\gn (0)$ is in agreement with that 
of $N$(0) reported in photoemission spectroscopy measurements on La214 
performed by Ino \textit{et al}\cite{Ino1}. We can also obtain the information about $N(0)$ 
from the $T$ dependence of the NMR Knight shift at $T 
\lesssim  \Tc $, because the Knight shift 
drops below $\Tc $ when the Fermi surface is removed by the formation of the 
spin-singlet pairing gap. The $p$ dependence of $N(0)$, obtained from the data of 
the Cu NMR Knight shift reported by Ohsugi \textit{et al}. for La214\cite{Ohsugi1,Matsuzaki2}, also 
agrees with that of $\gn (0)$, as shown in Fig. \ref{f5}
. The agreement 
between $\gn (0)$ and $N(0)$ gives validity for the present 
extrapolating process of $\gn $ $(T>\Tsf )$ down to below $\Tsf $. 
Thus we calculated the condensation energy $U(0)$ using the data of $\gamma 
_{s}(T)$ and the extrapolated $\gn $ $(T<\Tsf )$, and plotted the 
result against $x$ in Fig. \ref{f6}
.

As seen in Fig. \ref{f5}
, $\gn (0)$, namely $N(0)$, decreases rapidly at $x 
\lesssim 0.2$. The decreases of $N(0)$ at 
$x \lesssim 0.2$ result, in large part, from 
the downward shift of the flat band from $\EF $ near $(\pmax \pi ,0)$ and 
$(0,\pmax \pi )$, i.e. the large pseudogap, as was demonstrated by ARPES 
measurements on La214\cite{Ino2}. The flat band, one of the striking features 
of the electronic structure in high-$\Tc $ cuprates, makes a large 
contribution to DOS, and so the downward shift of the flat band from 
$\EF $ leads to a large reduction of $N(0)$. Furthermore, since the small 
pseudogap as well as the large pseudogap grows in La214 samples for $x 
\lesssim 0.2$\cite{Momono3}, as mentioned above, 
the development of the small pseudogap at $T<T'$ $(\sim \Tco )$ also contributes 
to the reduction of $N(0)$, i.e., $\gn (0)$ for $x 
\lesssim 0.2$. 

The SC condensation energy $U(0)$ can be given approximately by
\begin{equation}
\label{eq2}
U(0) \approx \frac{\alpha }{2}N(0)\Delta _0 ^2 \approx \frac{2.1\times 10^{ 
- 5}}{2}\alpha \gamma _n (0)\Delta _0 ^2 [\mbox{eV}]
\end{equation}

\noindent
for a $d$-wave superconductor ($\alpha  \simeq $0.4) with a nearly flat 
DOS\cite{Sun1}. The expression can be expected to hold good in any 
reasonable models for the $d$-wave superconductivity\cite{Anderson2}. We 
calculated the condensation energy $U(0)$ using eq. (\ref{eq2}) for $\gn (0)$ 
(J/K$^{2}$mole) (Fig. \ref{f5}
) and $\Delta _{0}$ (eV) determined in tunneling 
experiments on La214 (Fig. \ref{f3}
 )\cite{Nakano2,Oda1,Matsuzaki1}. For the $x=0.22$ 
sample, whose SC anomaly of $\Cel $ is of typical BCS type (the inset of 
Fig. \ref{f2}
), $\Delta _{0 }$was estimated from $\Tc $ using the BCS relation 
$2\Delta _{0}=4.3\kB \Tc $ for $d$-wave superconductors\cite{Won1}. In the 
present study, since the value of $\gs $ at $T=0$, estimated from the 
$\Cel /T$ vs. $T^{2}$ plot, shows a small residual value ($\go $) for 
$x \lesssim 0.14$, we calculated the 
condensation energy $U(0)$ for both $\gn (0)$ and $\gn ' = 
\gn (0) -\go $. The condensation energy $U(0)$ thus 
calculated is in good agreement with the experimental result for the 
overdoped $x=0.22$ sample, as seen in Fig. \ref{f6}
. On the other hand, the calculated 
value for $x \lesssim 0.2$ is quite different from the 
experimental result. 

The serious disagreement between experimental and calculated values of 
$U(0)$ at $x \lesssim 0.20$ seems to result from the 
modification of the nature of the pairing gap caused by the pseudogap 
formation. In fact, it has been pointed out that $\Tc $ roughly scales with 
$\kappa p\Delta _{0}$ ($\kappa \sim 1.7$), instead of $\Delta _{0}$, 
over a wide $p$ range in the pseudogap regime\cite{Ido1,Nakano2}, although a $d$-wave gap 
is completed over the entire original Fermi surface at $T \ll \Tc $, with the 
maximum value $\Delta _{0}$ at antinodal points near $(\pi ,0)$ and 
$(0,\pi )$. This indicates that the SC energy scale determining $\Tc $ will 
be proportional to $p\Delta _{0 }$in the pseudogap regime; namely, 
$\kB \Tc  \propto p\Delta _{0}$. The relation $\kB \Tc \sim 
p\Delta _{0 }$was phenomenologically predicted first by Lee and Wen for 
the underdoped spin-gap (small pseudogap) regime although they took $\Delta 
_{0}$ to be almost independent of $p$ there, and then discussed 
microscopically on the basis of the SU(\ref{eq2}) slave-boson model\cite{Lee4,Wen1}. 
Recently Tesanovic has also discussed the relation in terms of 
vortex-antivortex pairs\cite{Tesanovic1}. Thus we recalculated $U(0)$ by substituting 
the SC energy scale $\beta p \Delta _0$ ($\beta = 4.5$) for $\Delta _{0}$ in eq. (\ref{eq2}), 
and plot the calculated result in Fig. \ref{f6}
. The newly calculated values 
reproduce the experimental result of $U(0)$ very well over the whole $p$-range 
examined. 

The new SC energy scale $\beta p \Delta _0$ ($\beta = 4.5$), introduced in the above 
analysis for $U(0)$, becomes smaller than $\Delta _{0}$ at doping levels for 
the pseudogap regime ($p \lesssim  0.2$). However, it returns back 
to $\Delta _{0}$ around the doping level $p$(=$x)$=0.22, where no pseudogap 
behavior appears and the SC properties are of the BCS type. Furthermore if 
we represent the SC energy scale $\beta p \Delta _0$ ($\beta = 4.5$) as 
$\Deff $, the phenomenological relation 
$\kB \Tc \sim \kappa p\Delta _{0}$ ($\kappa \sim $1.7) can be 
rewritten as $2\Deff \sim 5.3\kB \Tc $, which is 
similar to the BCS result for a $d$-wave superconductor. These results suggest 
that $\Deff =\beta p \Delta _0$ ($\beta = 4.5$) may correspond to the 
maximum value of the effective SC gap in the pseudogap regime. In subsection 
\ref{3.5}, $\Deff $ will be discussed in terms of the shrinkage of 
the coherent part of the pairing gap. 

\subsection{Comparison of $U(0)$ between La214 and Other Systems}

Here we compare the present data of $U(0)$ with those reported by Loram's group 
on Y$_{0.8}$Ca$_{0.2}$Ba$_{2}$Cu$_{3}$O$_{6 + \delta }$ (Y123(Ca)) and 
Bi2212\cite{Loram1,Loram2,Loram3}. Loram's group estimated the condensation energy $U(0)$ on 
the assumption that the hypothetical normal value of $\gn $, 
$\gn (T<\Tsf )$, is strongly $T$-dependent in samples for $p 
\lesssim  0.19$ and reduced to zero at 
$T=0$\cite{Loram1,Loram2,Loram3}. This is because the $T$ dependence of $\gn $ in the normal state 
($T>\Tc )$ could be reproduced by assuming the existence of a $T$-independent 
gap structure near $\EF $ which would reduce $\gn $ to be 0 at 
$T=0$. However, since the existence of such a gap structure has 
not been clarified yet experimentally\cite{Renner2,Dipasupil1}, we presumed a simple 
$T$-dependence for $\gn $ $(T<\Tc )$ with a finite value at $T=0$, as 
mentioned in subsection 
\ref{3.3}. This discrepancy in the $T$ dependence of $\gn $ $(T<\Tc )$ may prevent us from comparing the present data with those 
of the Loram group. However, we can confirm phenomenologically that $U(0)$ is 
almost independent of the $T$-dependence of $\gn $ $(T<\Tc )$ as long as 
the entropy balance holds between $T$ dependences of $\gn $ and 
$\gs $ at $T \le \Tsf $. This allows us to make a comparison 
between the present data of $U(0)$ and the Loram group's, because the entropy 
balance is satisfied in both groups' analyses. 
\begin{figure}[htbp]
\begin{center}
\includegraphics[width=0.95\linewidth,clip]{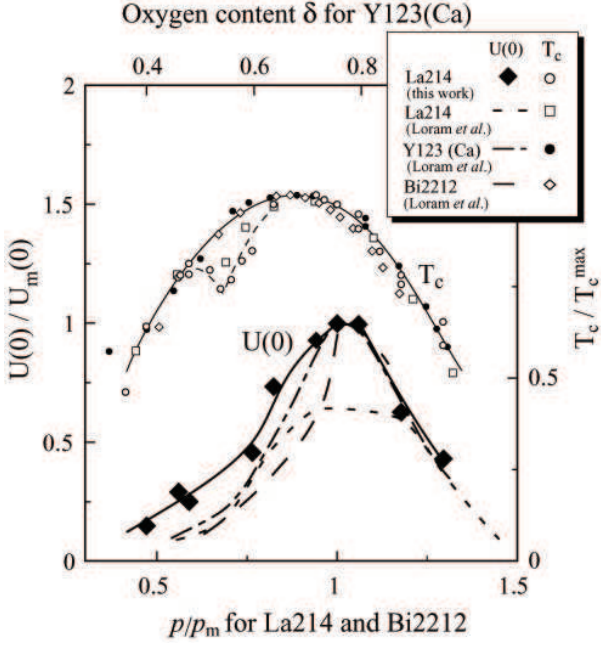}
\end{center}
\caption{Doping-level dependence of $U$(0) for La214, 
Y$_{0.8}$Ca$_{0.2}$Ba$_{2}$Cu$_{3}$O$_{6 + \delta}$(Y123(Ca)), 
Bi$_{2}$Sr$_{2}$CaCu$_{2}$O$_{8 + \delta }$(Bi2212), plotted 
together with $\Tc $. The $U(0)$ for Y123(Ca) and Bi2212 reported by Loram \textit{et al}. 
are shown by the dashed dotted line and dashed line, respectively.\cite{Loram1,Loram2,Loram3} $U(0)$ and $\Tc $ are normalized with their maximum values $\Um (0)$ and 
$\Tcmax $. The doping level $p$ and oxygen content $\delta $ are 
normalized with $\pmax $ and $\delta _{\rm m}$ where $U(0)$ takes the maximal 
value$ \Um (0)$. The $U(0)$ for La214 reported by Loram \textit{et al}. is also plotted with 
the same reduced scale (dotted line).\cite{Loram1}}
\label{f7}
\end{figure}

In Fig. \ref{f7}
, both $U(0)/\Um (0)$ and $\Tc /\Tcmax $, normalized with the 
maximum values of 

$U$ (0) and $\Tc $ respectively, are shown for La214 and Bi2212 as a function 
of $p/\pmax $, where $\pmax $ is the doping level at which $U(0)$ takes the maximal 
value $\Um (0)$. On the other hand, the original data of $U(0)$ and $\Tc $ for 
Y123 (Ca) are plotted against oxygen content $\delta $, instead of $p$, which 
has one-to-one correspondence with $p$. In Fig. \ref{f7}
, we convert $\delta $ into 
$p$ so that both peaks of $U(0)/\Um (0)$ vs. $\delta $ and 
$\Tc /\Tcmax $ vs. $\delta $ curves for Y123(Ca) should agree with 
the corresponding peaks for La214, respectively. The 
$\Tc /\Tcmax $ vs. $p/\pmax $ curves thus obtained for Y123(Ca) and 
Bi2212 are in agreement with that for La214 over a wide $p$ range except around 
$p=1/8$ in La214 and the Ortho-I ($\Tc \sim 60$K) phase in Y123 where 
$\Tc $ becomes less dependent on $p(\delta)$. The values of 
$U(0)/\Um (0)$ for Y123(Ca) and Bi2212 show $p$ dependences similar to that for 
La214, though $U(0)/\Um (0)$ tends to decrease faster at $p<\pmax $ in Y123(Ca) 
and Bi2212 than in La214. The similarity in both $U(0)$/$\Um $(0) vs. 
$p/\pmax $ and $\Tc /\Tcmax $ vs. $p/\pmax $ curves among La214, Y123(Ca) and 
Bi2212 implies that there exists no essential difference in the SC 
transition mechanism among them, though $\Tc $ varies from system to system. 

The condensation energy $U(0)$ reported by Loram's group for La214 is also 
plotted in Fig. \ref{f7}
, where $p$ is normalized so that the $\Tc $ vs. $p$ curves 
coincide with each other\cite{Loram1}. The overall $p$ ($=x$) dependence of $U(0)$ is 
qualitatively consistent, but the Loram data are smaller in the pseudogap 
regime ($x \lesssim  0.2$) than the present data. The 
difference of $U(0)$ can be attributed to the difference between the raw data 
of $\Cel $ measured by Loram's group and ours, because both groups adopted 
different ways to estimate the phonon part $\Cph $. One of the differences 
is that Loram's group adopted Zn impurity to destroy the superconductivity 
in the standard sample on the overdoped sample, whereas Ni impurity was used 
in the present study. 

\subsection{Coherent Pairing Gap}\label{3.5}

In high-$\Tc $ cuprates with low doping levels, the nodal parts of the Fermi 
line near $(\pi /2, \pi /2)$ dominate the in-plane transport; i.e., 
in-plane mobility of carriers on the nodal parts of the Fermi surface is 
much higher than on the antinodal parts near $(\pi ,0)$ and $(0,\pi )$\cite{Pines1,Geshkenbein1,Furukawa1}.
This is typically demonstrated by the following observation: 
the Fermi surface is truncated near $(\pi ,0)$ and $(0,\pi )$ by the small 
pseudogap formation at $T \lesssim  \Tco $, leaving the so-called 
nodal Fermi arcs\cite{Norman1}, but the in-plane resistivity remains almost 
unchanged\cite{Yanase1}. The pairing of m;;bile carriers with high in-plane 
mobility is expected to play a crucial role in making the collective pair 
motion coherent\cite{Pines1,Geshkenbein1}. Therefore;he carriers on the nodal Fermi 
arcs are expected to drive the SC phase transition when they start to form 
pairs\cite{Wen1,Pines1,Geshkenbein1,Furukawa1,Lauchli1}. This directs our attention to the 
possibility that in the pseudogap regime the pairing gap formed over the 
nodal Fermi arcs will function as the coherent pairing gap, namely, the 
effective SC gap which dominates $\Tc $ and $U(0)$. 
\begin{figure}[htbp]
\begin{center}
\includegraphics[width=0.95\linewidth,clip]{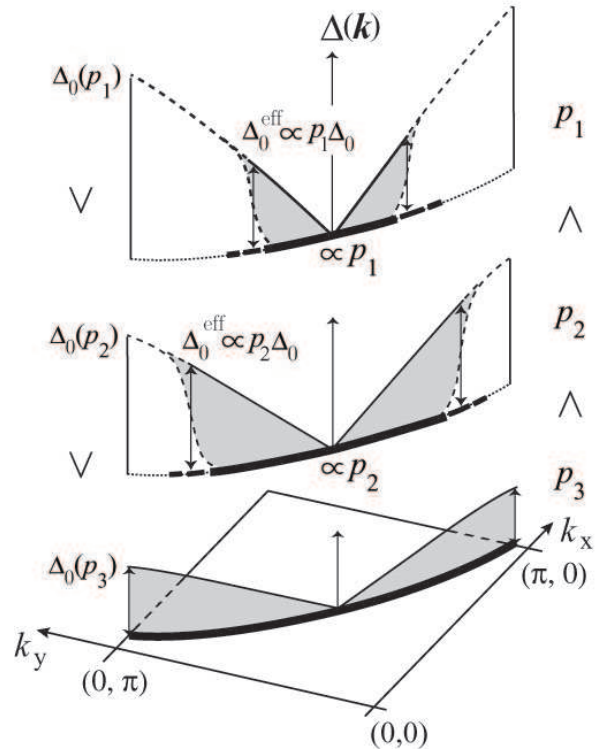}
\end{center}
\caption{Schematic illustration for doping dependence of a small pseudogap at 
$T\sim \Tc $ (dotted line) and the nodal Fermi arc (bold solid line) in 
$k_{x}-k_{y}$ space. The effective SC pairing gap (grayed area) is developed 
over the nodal Fermi arc at $T<\Tc $, and the effective SC gap together with 
the small pseudogap completes a $d$-wave energy gap over the underlying Fermi 
line. The upper, middle and bottom figures correspond to an underdoping 
level ($p_{1}$), nearly optimal doping ($p_{2}$), and the overdoping 
($p_{3}$) at which no pseudogap behavior appears, respectively. A line with 
arrow heads shows the maximum value of the effective SC gap $\Deff $ ($ \propto p\Delta _{0}$), where $\Delta _{0}$ is the 
maximum value of the energy gap at $T<\Tc $ ($\Delta _{0}(p_{1})>{\rm 
g}\Delta _{0}(p_{2})>{\rm g}\Delta _{0}(p_{3})$).}
\label{f8}
\end{figure}

The hypothetical normal state value $\gn (0)$, namely $N(0)$, will 
give us information about the linear dimension of the Fermi arc at $T\sim 
\Tc $, because it will be proportional to the dimension of the arc if the 
local DOS, $n(0)$, is constant over the Fermi line. As seen in Fig. \ref{f5}
, $\gn (0)$
 and $N(0)$ show a tendency for linear $p$ dependence below $x\sim 0.2$, 
suggesting that the dimension of the nodal Fermi arc will be roughly 
proportional to $p$ ($=x$) there. Furthermore, because of the distortion of the gap 
function caused by the higher harmonic in high-$\Tc $ cuprates, the pairing 
gap at $T \ll \Tc $ has a rather linear dispersion along the Fermi line in a 
slightly overdoped region as well as the underdoped one\cite{Mesot2}. Thus, 
if the dimension of the nodal Fermi arc at $T\sim \Tc $ is proportional to 
$p$ ($=x$), as suggested above, the gap size at the edge of the nodal arcs is 
roughly proportional to $p\Delta _{0}$, as shown schematically in Fig. \ref{f8}
. 
This gives an explanation for the maximum value of the effective SC gap 
$\Deff =\beta p\Delta _{0}$ ($\beta ={\rm const.}$), 
introduced in the present analysis for $U(0)$. Such a picture is almost the 
same as the scenario proposed by Wen \textit{et al}. to explain the relation 
$\kB \Tc \sim p\Delta _{0 }$microscopically on the basis of the SU(\ref{eq2}) 
slave-boson model, which predicts that a hole pocket with no shadow band at 
$\omega =0$ will appear near $(\pi /2, \pi /2)$, namely the nodal Fermi arc, in the 
spin gap regime\cite{Wen1,Lee3}. 

It is noteworthy here that the results of the electronic Raman scattering 
measurements on high-$\Tc $ cuprates are consistent with the present 
proposition that the coherent part of the pairing gap shrinks toward nodal 
points as the small pseudogap grows. Electronic Raman scattering 
measurements are expected to give more direct information about the coherent 
pairing gap than ARPES and tunneling spectroscopy measurements, because the 
electronic Raman response to superconductors is relevant to the coherence 
factor with even parity\cite{Opel1}. It has been reported for La214 as well 
as Bi2212 and Y123 that the B$_{\rm 1g}$ Raman continua, weighing out the 
antinodal parts of the Fermi surface, are indicative of marked suppression 
of the coherent pairing gap in the underdoped region, whereas the B$_{\rm 2g}$ 
continua, weighing out nodal parts, indicate the existence of the coherent 
pairing gap $\Dcoh $\cite{Opel1,Naeini1,Deutscher1,Momono5}. Such results in Raman 
scattering experiments mean that the coherent pairing gap will be located 
near nodal points, at least, in the underdoped region, which is consistent 
with the present proposition. The coherent pairing gap $\Dcoh $ may 
correspond to the effective SC gap determining $\Tc $ and $U(0)$. 

The present proposition is also consistent with recent ARPES results 
measured by Zhou \textit{et al}. on underdoped La214 samples; the well-defined 
quasiparticle peak exists on the nodal part of the Fermi line, but away from 
the nodal part it becomes broader and fades out rather abruptly\cite{Zhou1}. 

\section{Summary}

The electronic specific heat $\Cel $ was systematically measured on La$_{2 - 
x}$Sr$_{x}$CuO$_{4}$ (La214) to reexamine the development of the small 
pseudogap in the normal state ($T>\Tc )$ and the superconducting condensation 
energy at $T$=0,$ U$(0). The results obtained in the present study are summarized 
as follows.

1) The small pseudogap behavior appears in the $\gamma -T$ plot at 
temperature $T'\sim \Tco $; the $\gamma $ value, namely $N(0)$, shows a small 
bump at $T'$ and is progressively suppressed at $T<T'$. The temperature $T'$ $(\sim 
\Tco )$ roughly correlates with the onset temperature of the enhanced 
Nernst signal, reported by Wang \textit{et al}. on La214\cite{Wang1}. 

2) We confirmed that $U(0)$ was markedly reduced in the pseudogap regime. The 
reduction of $U(0)$ can be quantitatively explained by using the effective SC 
energy scale $\Deff =\beta p\Delta _{0}$ ($\beta =4.5$), 
instead of $\Delta _{0}$, and the DOS associated with the nodal Fermi 
surface which is removed by the pairing gap formation at $T<\Tc $. 

3) To explain the effective SC energy scale $\Deff $ in the 
pseudogap regime, we pointed out the possibility that the pairing gap formed 
over the nodal Fermi arcs at $T<\Tc $ will play a role as the coherent 
pairing gap, namely, the effective SC gap determining $\Tc $ and $U(0)$. 

\section*{Acknowledgment}
We would like to thank professor F. J. Ohkawa, professor S.-H. S. Salk and 
professor Z. Tesanovic for many stimulating discussions. This work was 
supported in part by Grants-in-Aid for Scientific Research on Priority Area 
(Novel Quantum Phenomena in Transition Metal Oxides) and on Projects (No 
13440104, 15340104, 15740193) from the Ministry of Education, Culture, 
Science, Sports and Technology of Japan.


\begin{thebibliography}{10}
\bibitem{Matsuzaki1}
The maximum gap $\Delta _{0 }$ for $x=0.1$ was measured by point-contact
  junction tunneling spectroscopy on single crystals in the present study.
\bibitem{Matsuzaki2}
The KS for $x=0.24$ reported in ref. 38 has a large residual value at $T \ll
  \Tc $, presumably, due to some inhomogeneous effect. Then we estimated
  $N$(0)$_{\rm KS}$ for $x=0.24$ by assuming that the KS for $x=0.24$ would drop to
  a level comparable with that for other doping levels at $T \ll \Tc $ if the
  inhomogeneous effect on the KS could be removed.
\end{thebibliography}
\end{document}